\documentclass[12pt,a4paper]{article}

\usepackage{amsmath}
\usepackage{mathtools}
\usepackage{caption}
\usepackage{subcaption}
\usepackage{indent first}
\usepackage{jheppub}
\usepackage{pdfcomment}
\usepackage{tikz}
\usetikzlibrary{shapes,matrix}
\usepackage{ytableau}

\def\Nequals#1{$\mathcal{N}{=}#1$}

\def\cT{\mathcal{T}}
\def\su{\mathfrak{su}}

\def\su{\mathfrak{su}}

\def\e{\mathfrak{e}}

\def\fg{\mathfrak{g}}
\def\ff{\mathfrak{f}}
\def\fsu{\mathfrak{su}}
\usepackage{framed}
\definecolor{shadecolor}{rgb}{0.90,0.90,0.90}
\newenvironment{claim}{\begin{shaded}\noindent\itshape\ignorespaces}{\end{shaded}}

\usepackage{xcolor}

\begin{document}

\title{$S^1/T^2$ Compactifications of 6d \Nequals{(1,0)} Theories and Brane Webs}
\abstract{
	We consider the circle and torus compactification of a certain subclass of 6d \Nequals{(1,0)} SCFTs which are Higgsable to the higher rank E-string theories.
		Using the T-duality between Type I' and Type IIB, we found that the $S^1$ compactification of the theories can be realized by 5-brane webs describing the 5d uplifting of a specified class $S$ theory, generalizing the result by Benini, Benvenuti and Tachikawa.
	We checked the above result by calculating conformal and flavor central charges of the 4d torus compactified theory both from the tensor branch structure of the 6d theory and from the predicted class $S$ description.
}
\author[1]{Kantaro Ohmori,}
\author[1]{Hiroyuki Shimizu,}
\affiliation[1]{Department of Physics, Faculty of Science, \\
 University of Tokyo,  Bunkyo-ku, Tokyo 133-0022, Japan}
\preprint{UT-15-36}

\maketitle

\section{Introduction and Summary}\label{sec:introduction}

The 6d \Nequals{(2,0)} theories have given us enormous information about lower dimensional supersymmetric field theories, for example \cite{Gaiotto:2009we}. Recently, the study of 6d \Nequals{(1,0)} superconformal field theories has gathered attention, expecting that those theories also shed light on a vast range of field theories.



Given the F-theory classification \cite{Heckman:2013pva,Heckman:2015bfa} for 6d SCFTs, the $S^1/T^2$ compactification of 6d \Nequals{(1,0)} theories is studied in \cite{Ohmori:2015pua, Ohmori:2015pia, DelZotto:2015rca}, 
generalizing the known relation between the 6d \Nequals{(2,0)} theories and 4d \Nequals{4} theories to the relation between the theories with eight supercharges.
It was shown there that the $S^1/T^2$ compactification of a ``very Higgsable" 6d \Nequals{(1,0)} theory, which means a theory flowing to the free hypermultiplets at a generic point of its Higgs branch, is a 5d/4d SCFT
at the origin of vacuum moduli without Wilson lines.

A typical example of very Higgsable 6d theories is the rank-$N$ E-string theory, which is the worldvolume theory on $N$ M5-branes probing the $\e_8$ wall in heterotic-M theory \cite{Ganor:1996mu}.
The $S^1$ compactification of rank-$N$ E-string theory results in the 5d SCFT with the $\e_8$ global symmetry. The 5-brane web constructions \cite{Aharony:1997ju, Aharony:1997bh} for these SCFTs are given in \cite{Benini:2009gi}, which are regraded as 5d uplifting certain class S theories. When $N=1$, this 4d \Nequals{2} SCFT is known to be the $\e_8$ Minahan-Nemeschansky theory \cite{Minahan:1996cj}.

\begin{figure}[ht]
	\centering
	\includegraphics[width=0.5\linewidth]{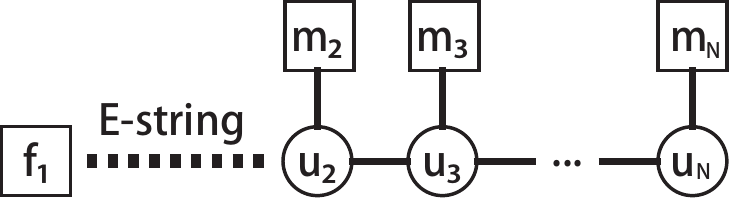}
	\caption{The tensor branch quiver of the theory $\cT^\text{6d}\{u_i\}$ which we consider in this paper.
		The $N-1$ positive integers $u_i$, $m_i$ for $i=2,\cdots,N$ specify the gauge and flaovr groups $\fg_i=\fsu(u_i)$, $\ff_i=\fsu(m_i)$ of the quiver.
		As usual, the solid lines represent the bifundamental hypers between connected gauge or flavor groups.
		The gauge group $\fsu(u_2)$ also couples to the rank 1 E-string theory represented by the dashed line.
		For each gauge node, there is also a tensor multiplet, whose vev determines the coupling of the gauge group on the node.
	}
	\label{fig:quiver}
\end{figure}

In this paper, we generalize the result of \cite{Benini:2009gi} to a class of 6d SCFTs with $N$ tensor modes whose tensor branch quivers are illustrated in Figure \ref{fig:quiver}.
For the readers familiar with the notation used in \cite{Heckman:2013pva,DelZotto:2014hpa}, this quiver can be represented as
\begin{equation}
	\begin{matrix}
		[\ff_1]& \varnothing &\fsu(u_2)&\fsu(u_3)& \cdots &\fsu(u_N)\\
		& 1 & 2 &2&\cdots & 2 
	\end{matrix}.
	\label{eq:tbranch}
\end{equation}
The leftmost tensor mode without gauge symmetry comes from the rank 1 E-string theory in Figure \ref{fig:quiver} on its tensor branch. When all $u_i$ are $0$, \eqref{eq:tbranch} is nothing but the tensor branch theory of the rank $N$ E-string theory.

The integers $u_i$ and $m_i$ which define the quiver are constrained by gauge anomally cancellation condition \cite{Ohmori:2014kda} so that they satisfy $m_i= 2u_i-u_{i+1}-u_{i-1}$ for $i=2\cdots N$ with $u_1=u_N=0$.
Further, the fact that $\fsu(u_2)$ should be a subgroup of the $\mathfrak{e}_8$ flavor symmetry of the rank 1 E-string theory requires $u_2\le 9$.
The flavor symmtery $\ff_1$ in the quiver of Figure \ref{fig:quiver} is the commutant of $\fsu(u_2)$ embedded in $\mathfrak{e}_8$.
 
Let us introduce a notation which specifies the theories we consider.
For $u_2\le 7$ case we denote the 6d theory \eqref{eq:tbranch} by $\cT^\text{6d}\{u_i\}$.
When $u_2=8$, there are two choices of the flavor; $\ff_1=\su(2)$ and $\ff_1=\mathfrak{u}(1)$, and we denote the former theory as $\cT^\text{6d}\{u_i\}$ and the latter theory as $\cT_*^{\text{6d}}\{u_i\}$.
We also use the symbol $\cT_*^\text{6d}\{u_i\}$ in the case where $u_2=9$. We denote the $S^1/T^2$ compactification of the 6d theory $\cT_{(*)}^\text{6d}\{u_i\}$ as  $\cT_{(*)}^\text{5d/4d}\{u_i\}$.


Our main claim for the $S^1/T^2$ compactification is 
\begin{claim}
	\begin{equation}
		\cT^\text{5d}\{u_i\} = \widehat{\mathsf{T}}_{K}\{Y_1,Y_2,Y_3\}, \label{eq:result1}
	\end{equation}
	where $\widehat{\mathsf{T}}_{K}\{Y_1,Y_2,Y_3\}$ is the 5d uplifting of the 4d Class $S$ theory $\mathsf{T}_{K}\{Y_1,Y_2,Y_3\}$ of type $A_{K}$, whose UV curve is the sphere with three punctures $Y_1$, $Y_2,$ and $Y_3$.
	\footnotemark

	$K$ denotes $6N+n_7+n_8$, where $n_I = \#\{i=2,3,\cdots,N|u_{i+1}-u_i\ge I\}$.
	$Y_2$ and $Y_3$ are the partitions of $K$ defined by $Y_2=[2N+n_7+n_8,2N,2N]$ and $Y_3=[3N+n_7,3N+n_8]$.
	Let $Y_1^\text{T}=[\ell_1,\cdots,\ell_N]$ be the partition of $K$ obtained by taking the transpose of the Young diagram $Y_1$, then
	\begin{equation}
	\begin{cases}
		\ell_i =0  &(i\ge N-n_6+1)\\
		\ell_{N-i+2} = 6-u_{i}+u_{i-1} & (i=2,\cdots,N-n_6)\\
		\ell_1 = 6+u_N.
	\end{cases}
	\label{eq:Y1def}
	\end{equation}
	
	The 4d version of the statement 
	\begin{equation}
		\cT^\text{4d}\{u_i\} = \mathsf{T}_{K}\{Y_1,Y_2,Y_3\}
		\label{eq:claim4d}
	\end{equation}
	automatically follows.
\end{claim}
	\footnotetext{ The notation here is adopted from \cite{Tachikawa:2015bga}.}
When $u_i=0$ for all $i=2,\cdots,N$, $\cT^\text{6d}\{u_i=0\}$ is the rank $N$ E-string theory, and the corresponding class S theory is $\widehat{\mathsf{T}}_{6N}\{[N^6],[2N,2N,2N],[3N,3N]\}$ which is proposed in \cite{Benini:2009gi} as the $S^1$ compactification of the rank $N$ E-string theory.
\footnote{ When $u_i=1$ for $i=2,\cdots,N$, $\cT^\text{6d}\{u_i=1\}$ is the rank $N$ E-string theory plus a decoupled hyper, and the corresponding theory is $\widehat{\mathsf{T}}_{6N}\{[N^5,N-1,1],[2N,2N,2N],[3N,3N]\}$, which was firstly observed by the index calculation \cite{Gaiotto:2012uq}.}
Thus our claim generalizes the result of them. For the compactifications of $\cT^\text{6d}_*\{u_i\}$, the claim is 
\begin{claim}
	\begin{align}
		\cT^\text{5d}_*\{u_i\} &= \widehat{\mathsf{T}}_{K_*}\{Y_1,Y_2^*,Y_3^*\},\label{eq:result2}\\
		\cT^\text{4d}_*\{u_i\} &= \mathsf{T}_{K_*}\{Y_1,Y_2^*,Y_3^*\}, \label{eq:claim4d2}
	\end{align}
	where $K_*=6N+n_7+n_8+n_9$, $Y_2^*=[2N+n_7,2N+n_8,2N+n_9]$, and $Y_3^*=[3N+n_7+n_8+n_9,3N]$.
	$Y_1$ is defined by the same equations as the former case.
	When $u_2\le7$, $K_*=K,Y_2^*=Y_2$ and $Y_3^*=Y_3$ holds.
\end{claim}
Note that a single 4d SCFT might admit multiple class S constructions, thus the above class S descriptions are not necessarily unique.

The organization of the rest of the paper is as follows. In section \ref{sec:setup},we read off the Type I' brane web construction of the 6d theory and use the Hanany-Witten transitions to take the brane system to the most convenient form for taking the T-dual.

In section \ref{sec:IIB}, by T-dualizing the Type I' brane construction, we will find the 5-brane web describing the 5d SCFT obtained by the $S^1$ compactification. The resulting web has three external legs of 5-branes terminated at 7-branes \cite{Benini:2009gi}, thus we will show the results \eqref{eq:result1} and \eqref{eq:result2}. Then, it follows that the $T^2$ compactification is given by the A-type 6d \Nequals{(2,0)} theory on a sphere with three punctures, confirming \eqref{eq:claim4d} and \eqref{eq:claim4d2}. 


In section \ref{sec:anomalies}, we will provide further evidence for the 4d version of our main claims \eqref{eq:claim4d} and \eqref{eq:claim4d2} by calculating 4d conformal and flavor central charges in two ways. 
First by the method of \cite{Ohmori:2015pua}, the charges are obtained from the 6d tensor branch structure, and then we get the same quantities from the corresponding class S description by using the methods developed in \cite{Chacaltana:2010ks,Chacaltana:2012zy} 

In section \ref{sec:conclusions}, we conclude this paper with a short discussion about possible future directions.

\paragraph{Note added:} During the final stage of the preparation of this paper, the works \cite{Zafrir:2015rga,Hayashi:2015zka} appeared on arXiv, that have a significant overlap with our results.

\section{Brane engineering in Type I'}\label{sec:setup}
\begin{figure}
	\centering
	\includegraphics[width=0.8\linewidth]{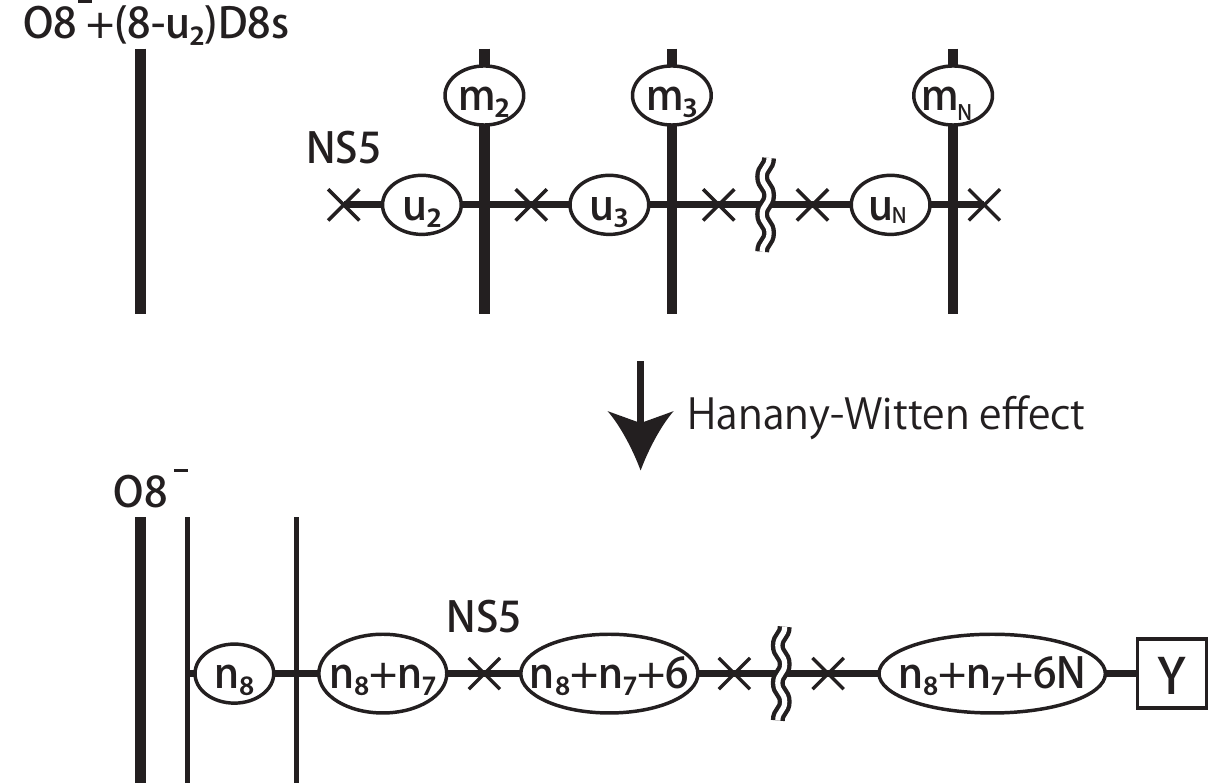}
	\caption{Upper: straighforward Type I' brane engineering \cite{Brunner:1997gk, Hanany:1997gh}. The $\times$ mark represents an NS 5 brane, the horizontal line represents the stack of D6 branes, and the vertical lines represent D8 branes or the stack of O8${}^-$ plane and D8 branes.
		The symbols in the circles are the numbers of the branes there.
	Lower: Type I' configuration after the Hanany-Witten transitions.
There are two D8 branes near the O8${}^-$ plane, each has $n_7$ and $n_8$ D6 branes ending on it, and $u_N+6$ D8 branes on the right side of the $N$th NS5 brane. The $K=n_8+n_7+6N$ D6 branes end on the stack of $u_N+6$ D8 branes, and the pattern of the ending is specified by the Young diagram $Y_1$ \eqref{eq:Y1def} \cite{Gaiotto:2014lca}.   }
	\label{fig:HZHW}
\end{figure}

In this section we describe the 6d theories $\mathcal{T}^{6d}\{u_i\}$ of our interest from brane engineering in Type I' \cite{Brunner:1997gk, Hanany:1997gh} (see also \cite{Brunner:1997gf, Gorbatov:2001pw, Gaiotto:2014lca}). It is pictorially shown in the upper half of Fig. \ref{fig:HZHW}.  We put an O8${}^-$ plane and D8 branes on $x^0\sim x^5, x^7\sim x^9$, D6 branes on $x^0\sim x^6$ and NS5 branes on $x^0\sim x^5$. Then the 6d theory is realized on $x^0 \sim x^5$.
\footnote{ In this paper we put only one O8${}^-$, thus actually consider Type IIA on $\mathbb{R}/\mathbb{Z}_2$ and always take the decoupling limit of gravity. We put $8+u_N$ D8 branes in total, which is prohibited in the case of full Type I' where there are two O8${}^-$ planes.}

We start from $N$ NS5 branes separated along $x^6$ direction and $u_i$ D6 branes suspended between $i$th and $(i-1)$th NS5 brane. There are $m_i \equiv 2u_i-u_{i-1}-u_{i+1}$ D8 branes intersecting with the $i$th bunch of D6 branes as required from the charge conservation in Type I' brane construction \cite{Brunner:1997gk, Hanany:1997gh}. Since the Romans mass at the O8${}^-$ plane is $-8$, there should be $8-u_2$ D8 branes near the O8${}^-$ plane to obtain the gauge group $\su(u_2)$ on the first bunch of D6 branes.

We should note that this setup reproduces the flavor symmetry $[\ff_i]$ in \eqref{eq:tbranch}. The flavor symmetry $\ff_i=\su(m_i)$ for $i\ge 2$ is realized by the $m_i$ D8-branes intersecting with the $i$th bunch of D6-branes. The flavor symmetry $\ff_1$ of $\cT^\text{6d}\{u_i\}$ is realized by the stack of the O8${}^-$ plane and D8 branes at $x^6=0$ when the dilaton background diverges at the O8${}^-$ plane.


In order to make the discussion in Sec \ref{sec:IIB} easier, we make the Hanany-Witten transitions by moving D8 branes, obtaining the Young diagram in \eqref{eq:Y1def} as the boundary condition of the D6-branes at the D8-branes. The resulting configuration is such that exactly two D8 branes are near the O8${}^-$ plane and all the other D8 branes are located where $x^6$ is very large and far away from the NS5 branes, as illustrated in the lower half of Fig \ref{fig:HZHW}.


The concrete procedure for the Hanany-Witten transitions is as follows. We move the leftmost and the second leftmost D8 branes in the upper half of Fig \ref{fig:HZHW} on top of the O8${}^-$ plane. \footnote{We can choose other D8 branes to be put on the O8${}^-$ plane, which leads other class S descriptions after the T-dual.}  In this process, the Hanany-Witten effect creates additional $n_7 = \#\{i=2,3,\cdots,N|u_{i+1}-u_i\ge 7\}$ and $n_8 = \#\{i=2,3,\cdots,N|u_{i+1}-u_i\ge 8\}$ D6 branes connecting the two D8 branes respectively and the first NS5 brane.

Then, we move the remain $6+u_N$ D8 branes near the NS5 stack to the region where $x^6$ is very large and far away from all the NS5 branes. The Hanany-Witten effect again creates additional D6 branes.  The Romans mass in the region where all the NS5 branes sit is now $-6$ and there are $6i+n_7 +n_8$ D6 branes suspended between $i$th and $(i+1)$th NS5 brane. 

There are also $6N+n_7+n_8$ D6 branes extending from the $N$th NS5 brane toward the region where $x^6$ is large.
These D6 branes end on the $6+u_N$ D8 branes. The ending pattern of the D6 branes to the D8 branes specifies the partition \eqref{eq:Y1def} \cite{Gaiotto:2014lca}. Namely, there are $\ell_i -\ell_{i+1}$ D8 branes on which $i$ D6 branes have end.
These D8 branes are aligned so as to satisfy the ordering constraint in \cite{Gaiotto:2014lca},
which says that the number of D6 branes ending on each D8 brane is a decreasing function as we move away from the region where all the NS5 branes sit. 

\paragraph{$\cT^\text{6d}_*\{u_i\}$ and O8${}^*$ plane.}

In the discussion so far, we use the O8${}^-$ plane in the brane construction. However, we can have an alternative orientifold 8-plane in Type I' brane engineering: O8${}^*$ plane \cite{Morrison:1996xf,Douglas:1996xp,Gorbatov:2001pw}.

In \cite{Morrison:1996xf,Douglas:1996xp}, the theory of a D4 brane probing the stack of O8${}^-$ plane and $n\le8$ D8 branes was investigated. When the dilaton background at O8${}^-$ diverges, the theory has $\mathfrak{e}_{n+1}$ flavor symmetry and called $E_{n+1}$ theory. Moreover, it was found that the $E_2$ theory has two distinct mass deformations which keep the diaton background infinite; one is called $E_1$ theory with $\mathfrak{e}_1=\su(2)$ flavor symmetry and another is called $\tilde{E}_1$ theory with $\tilde{\mathfrak{e}}_1=\mathfrak{u}(1)$ symmetry.
The $\tilde{E}_1$ theory has further mass deformation to the $E_0$ theory which has no flavor symmetry.

This indicates that there are two distinct ways of splitting one D8 brane out of the stack of O8${}^-$ plane and one D8 brane. They are realized using the different kind of orientifold 8-plane called O8${}^*$ in \cite{Gorbatov:2001pw} as follows:
\begin{equation}
	\begin{split}
	\text{O8${}^-$ $+$ D8} &\rightarrow \text{O8${}^-$},\text{D8}\\
			 &\stackrel{\searrow}{} \text{O8${}^*$}+\text{D8},\text{D8}\rightarrow \text{O8${}^*$},\text{D8},\text{D8}.
	\end{split}
	\label{eq:O8star}
\end{equation}
Here $+$ denotes the stack of two objects, while a comma means that the two objects exist separately.
As a consequence, the flavor symmetry $\ff_1$ associated to the O8${}^-$ plane with the divergent dilaton background is $\mathfrak{e}_1$, while that for O8${}^*+$D8 is $\tilde{\mathfrak{e}}_1$.

For the brane engineering of the theory $\cT^\text{6d}_*\{u_i\}$, we need to make use of the $O8^*$ plane. We follow the same procedure outlined above. First, we put O8${}^*$-D8-D6-NS5 to reproduce the configuration \eqref{eq:tbranch}. Note that in this case we take the Romans mass at the O8${}^*$ plane to be $-9$ and we put $9-u_2$ D8 branes near the O8${}^*$ plane. Second, we shuffle the D8 branes to obtain a configuration with three D8 branes near the $O8^*$ plane, having $n_{7,8,9}$ D6 branes ending each of them and the first NS5 brane. Finally, we move other D8 branes near the NS5 stack to the region where $x^6$ is very large and far away from all the NS5 branes. The Romans mass in the region where all the NS5 branes sit is again $-6$ and the number of D6 branes in the $i$th bunch is $6i + n_7 + n_8 +n_9$. The ending pattern of the D6 branes on the D8 branes where $x^6$ is large is again specified by the partition \eqref{eq:Y1def}.

\section{IIB web diagrams}\label{sec:IIB}
\begin{figure}
	\centering
	\includegraphics[width=0.35\linewidth]{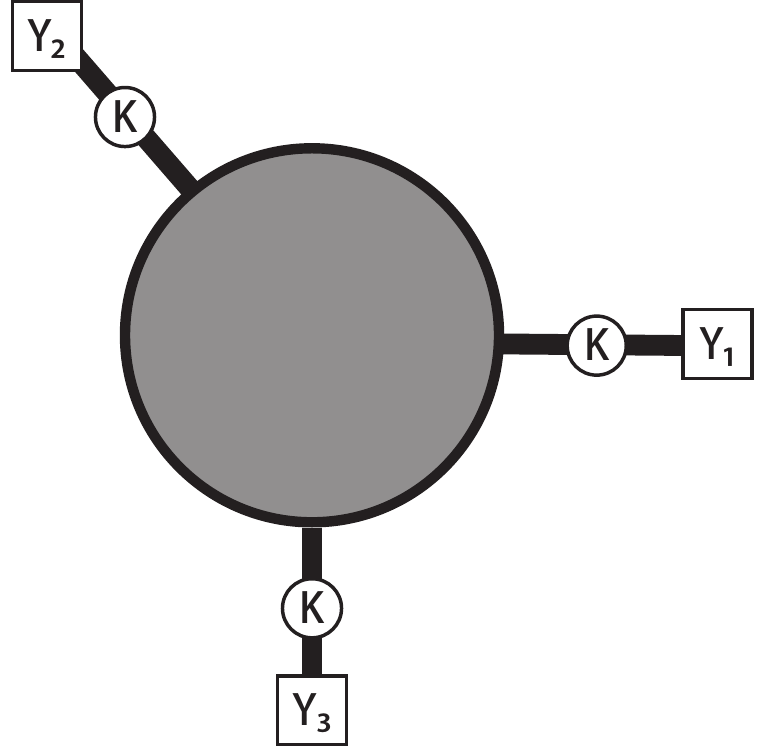}
	\caption{The 5-brane web configuration introduced in \cite{Benini:2009gi}. It has three legs made up of $K$ 5-branes of type $(1,0)$, $(0,1)$ and $(1,-1)$ respectively. The 5-branes in each leg terminate on 7-branes of the same type. The ending pattern of each leg at the 7-branes determines the Young diagram $Y_i $. Since the internal 5-brane web configuration is determined (up to flop transitions) by the boundary data $K$ and $Y_i \; (i=1,2,3)$, we do not write it explicitly. The 5d SCFT from this web is the 5d uplift $\widehat{\mathsf{T}}_{K}\{Y_1,Y_2,Y_3\}$ of the class S theory $\mathsf{T}_{K}\{Y_1,Y_2,Y_3\}$.}
		\label{fig:5dfixture}
\end{figure}

In this section, we establish the dualities \eqref{eq:result1}, \eqref{eq:claim4d}, \eqref{eq:result2} and \eqref{eq:claim4d2}. First of all, we briefly recall a class of 5d SCFTs introduced in \cite{Benini:2009gi} as 5d uplifts of some class S theories. Each of them is engineered by a junction of 5-branes with three legs which consist of $K$ 5-branes with charges $(1,0)$, $(0,1)$ and $(1,-1)$ respectively, as illustrated in Figure \ref{fig:5dfixture}. They are terminated at 7-branes of type $(1,0)$, $(0,1)$ and $(1,-1)$, respectively. The ending pattern of the 5-branes at the 7-branes specifies a partition of $K$ and then we associate a Young diagram $Y_i \; (i=1,2,3)$ for each leg. 

When we shrink the internal part of the web to a single point, we obtain the 5d SCFT $\widehat{\mathsf{T}}_{K}\{Y_1,Y_2,Y_3\}$, the right hand side of \eqref{eq:result1}. 
Upon further reduction to 4d, this 5d theory becomes the class S theory $\mathsf{T}_{K}\{Y_1,Y_2,Y_3\}$ in \eqref{eq:claim4d}.

To connect this 5-brane web construction of the 5d SCFT $\widehat{\mathsf{T}}_{K}\{Y_1,Y_2,Y_3\}$ with the Type I' brane engineering in Sec \ref{sec:setup}, we utilize T-duality and Hanany-Witten effect. This proceeds as follows. First, we T-dualize the Type I' brane configuration in Sec \ref{sec:setup} to obtain the Type IIB brane configuration with 5-branes and 7-branes, which corresponds to the $S^1$ compactification of $\cT^{\text{6d}}_{(*)}\{u_i\}$. 
Second, by taking a mass decoupling limit, we find the web configuration which describes the 5d SCFT $\mathcal{T}^{5d}_{(*)}\{u_i\}$ obtained by the zero radius limit $R_6 \to 0$. This mass deformation is achieved by moving one 7-brane toward the infinity without creating 5-branes due to Hanany-Witten effect.

Finally, we move the remaining 7-branes toward the infinity. During the process, Hanany-Witten effect creates additional 5-branes. We find that the resulting 5-brane web configuration is that of Figure \ref{fig:5dfixture}, a three pronged junction of 5-branes terminated at 7-branes. Thus, we establish the results \eqref{eq:result1}, \eqref{eq:claim4d}, \eqref{eq:result2} and \eqref{eq:claim4d2}. In the rest of this section, we explain the strategy outlined above more concretely.


\subsection{Notations on 7-branes}\label{sec:7brane}
Before moving to the concrete process, we summarize notations and conventions we use in the rest of this section about 7-branes in Type IIB \cite{Aharony:1997bh,DeWolfe:1998eu,DeWolfe:1998yf,Benini:2009gi}.
Let $\mathbf{X}_{[P,Q]}$ denotes the 7-brane with charge $[P,Q]$ where $P,Q$ are coprime. 
We use the following aliases $\mathbf{A}=\mathbf{X}_{[1,0]}, \mathbf{B}=\mathbf{X}_{[1,-1]}, \mathbf{C}=\mathbf{X}_{[1,1]}$, and $\mathbf{N}=\mathbf{X}_{[0,1]}$.
The monodromy matrix $K(\mathbf{X}_{[P,Q]})=K_{[P,Q]}$ of the 7-brane $\mathbf{X}_{[P,Q]}$ is
\begin{equation}
	K_{[P,Q]}=
	\begin{pmatrix}
		1+PQ & -P^2\\
		Q^2 & 1-PQ
	\end{pmatrix}.
\end{equation}
A 5-brane with charge $(p,q)$, when anti-clockwise crossing the branch cut of the 7-brane $\mathbf{X}_{[P,Q]}$, becomes a $(p',q')$ 5-brane where
\begin{equation}
	\begin{pmatrix} p' \\q'\end{pmatrix}
		=
		K_{[P,Q]}
	\begin{pmatrix} p \\q\end{pmatrix}
		=
	\begin{pmatrix} p \\q\end{pmatrix}
		-
		(Pq-Qp)
	\begin{pmatrix} P \\Q\end{pmatrix}.
		\label{eq:monod}
\end{equation}

\begin{figure}
	\centering
	\includegraphics[width=.75\linewidth]{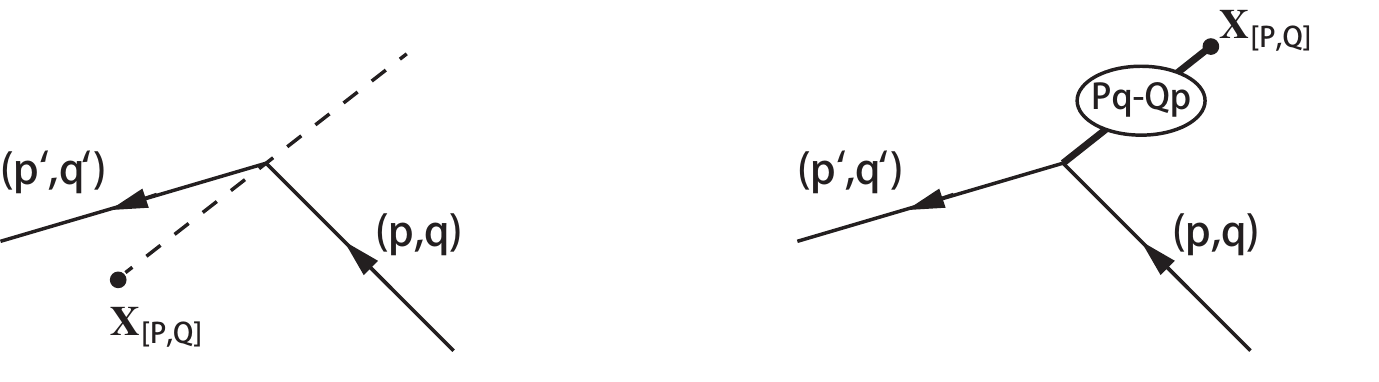}
	\caption{The Hanany-Witten effect between a 7-brane and a 5-brane.}
	\label{fig:HW}
\end{figure}

When a 7-brane $\mathbf{X}_{[P,Q]}$ crosses a $(p,q)$ 5-brane as in the Figure \ref{fig:HW}, the Hanany-Witten effect attaches $(P,Q)$ 5-branes to the 7-brane. 
The number of the emergent $(P,Q)$ 5-branes should be $|Pq-Qp|$ so that the tension balances at the trivalent point.

When there are some 7-branes $\mathbf{X}_{[P_1,Q_1]},\mathbf{X}_{[P_2,Q_2]},\cdots,\mathbf{X}_{[P_n,Q_n]}$ arranged anti-clockwise in this ordering, we denote the configuration by just writing them as
\begin{equation}
	\mathbf{X}_{[P_1,Q_1]}\mathbf{X}_{[P_2,Q_2]}\cdots\mathbf{X}_{[P_n,Q_n]},
\end{equation}
and the corresponding monodromy matrix as
\begin{equation}
	K(\mathbf{X}_{[P_1,Q_1]}\mathbf{X}_{[P_2,Q_2]}\cdots\mathbf{X}_{[P_n,Q_n]})
	=K_{[P_n,Q_n]}K_{[P_{n-1},Q_{n-1}]}\cdots K_{[P_1,Q_1]}.
\end{equation}
We can rearrange two 7-branes $\mathbf{X}_{[P_1,Q_1]},\mathbf{X}_{[P_2,Q_2]}$ by the following rule:
\begin{equation}
	\mathbf{X}_{[P_1,Q_1]}\mathbf{X}_{[P_2,Q_2]}=\mathbf{X}_{[P_2,Q_2]}\mathbf{X}_{[P'_1,Q'_1]}=\mathbf{X}_{[P_2',Q_2']}\mathbf{X}_{[P_1,Q_1]},
	\label{eq:7rearange}
\end{equation}
where
\begin{equation}
	\begin{pmatrix} P'_1 \\Q'_1\end{pmatrix}
		=
		K_{[P_2,Q_2]}
	\begin{pmatrix} P_1 \\Q_1\end{pmatrix}
		,\qquad
	\begin{pmatrix} P'_2 \\Q'_2\end{pmatrix}
		=
		K_{[P_1,Q_1]}
	\begin{pmatrix} P_2 \\Q_2\end{pmatrix}.\label{eq:change}
\end{equation}

We name some important 7-brane configurations such as
\begin{align}
	\mathbf{E}_N &= \mathbf{A}^{N-1}\mathbf{B}\mathbf{C}\mathbf{C} = \mathbf{A}^N \mathbf{X}_{[3,-1]} \mathbf{N},\\
	\widehat{\mathbf{E}}_N &= \mathbf{E}_N \mathbf{X}_{[3,1]} = \mathbf{A}^{N-1}\mathbf{B}\mathbf{C}\mathbf{B}\mathbf{C} = \mathbf{A}^{N}\mathbf{B}\mathbf{X}_{[1,2]}\mathbf{X}_{[2,1]}.
\end{align}
Here we assume that $N \geq 2$. When $N=1$, we cannot equate $\mathbf{E}_1=\mathbf{B}\mathbf{C}\mathbf{C}$ to $\mathbf{A}\mathbf{X}_{[3,-1]}\mathbf{N}$ by the operations \eqref{eq:7rearange} therefore the latter is an inequivalent configuration which is denoted as $\tilde{\mathbf{E}}_1$. We define $\mathbf{E}_0$ by $\mathbf{X}_{[3,-1]}\mathbf{N}$. The configuration $\widehat{\tilde{\mathbf{E}}}_1$ and $\widehat{\mathbf{E}}_0$ is again given by $\tilde{\mathbf{E}}_1\mathbf{X}_{[3,1]}$ and $\mathbf{E}_0 \mathbf{X}_{[3,1]}$ respectively.

\subsection{Warm up: T-dual of E-string theory}\label{sec:BBT}

\begin{figure}
	\centering
	\includegraphics[width=0.8\linewidth]{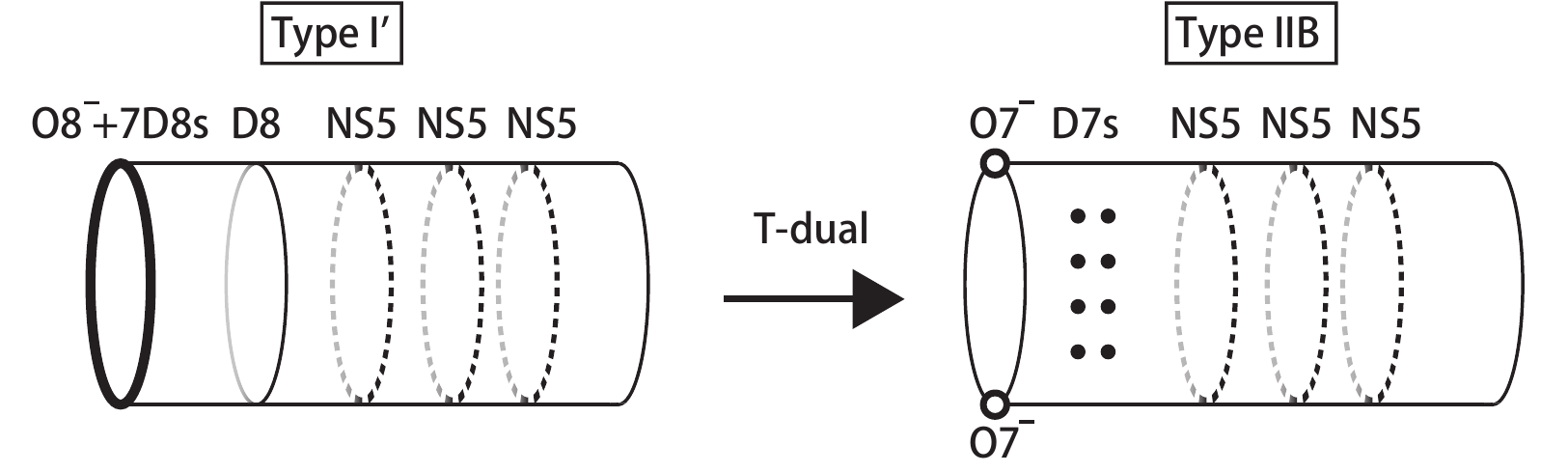}
	\caption{T-dual of the Type I' brane configuration realizing $S^1$ compactified higher rank E-string theory. The O8${}^-$ plane wrapping $S^1$ becomes two O7${}^-$ planes and the eight D8s become eight D7 branes, while the NS5 branes in type I' remain to be NS5.}
	\label{fig:est}
\end{figure}

To begin with, we start from the case where all the gauge algebras are empty in \eqref{eq:tbranch}, where the 6d theory is now the rank-$N$ E-string theory. While the result of this section was first obtained in \cite{Benini:2009gi}, we adopt the T-duality argument from \cite{Hayashi:2015fsa}. 
 
We start from the Type I' brane configuration where we have seven D8 branes on top of the O8${}^-$ plane and one D8 brane slightly away from the O8${}^-$ plane. There are also $N$ NS5 branes away from that O8${}^-$-D8 system where the Romans mass is $0$.

After the $S^1$ compactification, we can take the T-dual of the brane system to obtain the Type IIB O7${}^-$-D7-NS5 system, as illusrated in Figure \ref{fig:est}. Note that this T-dual is valid because in the Type I' configuration, the Romans mass is $0$ far from the O8${}^-$ plane, thus the dual Type IIB geometry should asymptotically be the cylinder.

\begin{figure}
	\centering
	\includegraphics[width=0.3\linewidth]{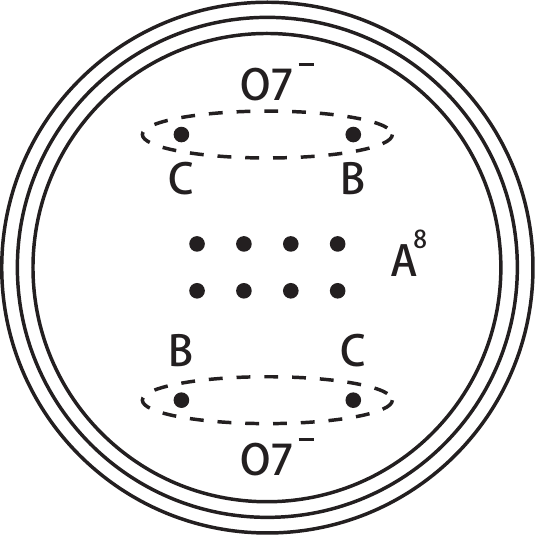}
	\caption{The Type IIB brane configuration in Figure \ref{fig:est} seen from the left.
	The O7${}^-$ planes splits into $\mathbf{B}$ and $\mathbf{C}$ branes, therefore there are twelve 7-branes wrapped by the $N$ circles of 5-branes.}
	\label{fig:BBT}
\end{figure}

Since the O7${}^-$ plane is the bound state of two 7-branes of type $\mathbf{B}$ and $\mathbf{C}$ \cite{Sen:1996vd} and the D7 brane is of type $\mathbf{A}$,
the system is equivalent to $N$ 5-branes encircling twelve 7-branes $\widehat{\mathbf{E}}_9=\mathbf{A}^8\mathbf{B}\mathbf{C}\mathbf{B}\mathbf{C}$ as shown in Figure \ref{fig:BBT}, which is considered in \cite{Kim:2015jba}.
Note that since each 7-brane has deficit angle $\frac16\pi$, the total deficit angle of twelve 7-branes is $2\pi$, 
thus the metric of the diagram Figure \ref{fig:BBT} is that of the cylinder outside of where 7 branes sit.
The same fact is also related to the fact $K(\widehat{\mathbf{E}}_9)=1$.

\paragraph{Mass decoupling of Kaluza-Klein modes.} The configuration in Figure \ref{fig:BBT} engineers the theory with Kaluza-Klein modes \cite{Kim:2015jba}. To obtain the 5d SCFT with $\mathfrak{e}_8 \times \mathfrak{su}(2)$ global symmetry from the E-string theory on $S^1$, we need to decouple the Kaluza-Klein modes by taking $R^6 \to 0$ preserving the global symmetry.  

This can be achieved by rearranging the 7-branes by $\mathbf{B}\mathbf{C}\mathbf{B}\mathbf{C}=\mathbf{B}\mathbf{C}\mathbf{C}\mathbf{X}_{[3,1]}$ and moving $\mathbf{X}_{[3,1]}$ toward the infinity, leaving the $\mathbf{E}_9$ 7-brane inside the circles of 5-branes. Here we show that we can make this decoupling without introducing additional 5-branes coming from the Hanany-Witten effect. 

To this end, we note that each 7-brane inside the circle has a branch cut that runs toward the infinity. When the circle of 5-brane crosses the cut, the $(p,q)$ charge of the 5-brane which makes up the circle changes to $(p',q')$ according to the formula \eqref{eq:change}. The fact $K(\widehat{\mathbf{E}}_9)=1$ ensures that the charge of the 5-brane comes back to its original value after crossing all the cuts from the 7-branes, as required by the consistency. We can choose the charge at a small segment in the circle to be $(3,1)$. Then, we can move the 7-brane $\mathbf{X}_{[3,1]}$ to the infinity through that segment without Hanany-Witten effect.

\paragraph{Pulling out 7-branes.} In order to obtain the 5-brane web as in Figure \ref{fig:5dfixture}, we rearrange the 7-branes and pull them out from the circles. We rearrange the five 7-branes $\mathbf{E}_3=\mathbf{A}^2\mathbf{B}\mathbf{C}\mathbf{C}$ in the remaining 7-branes $\mathbf{E}_9$ inside the circles as
\begin{equation}
\mathbf{E}_3 = \mathbf{A}^2 \mathbf{B}\mathbf{C}\mathbf{C} = \mathbf{B}\mathbf{N}^2 \mathbf{C}^2 = \mathbf{B}\mathbf{N}\mathbf{A}^2 \mathbf{N}= \mathbf{B}^3\mathbf{N}^2,\label{eq:rearrange} 
\end{equation}
where we used $\mathbf{A}\mathbf{B}=\mathbf{B}\mathbf{N}, \mathbf{N}\mathbf{C}=\mathbf{A}\mathbf{N}$ and $\mathbf{N}\mathbf{A}=\mathbf{B}\mathbf{N}$. Note that this rearrangement is nothing but moving two $\mathbf{A}$ branes from the leftmost to the rightmost in $\mathbf{E}_3$.

Then, we move the three types of 7-branes $\mathbf{A}$, $\mathbf{B}$ and $\mathbf{N}$ toward the infinity. To count the number of additional 5-branes created by Hanany-Witten effect, we concretely keep track of the charges of the circle of 5-brane. When decoupling the 7-brane $\mathbf{X}_{[3,1]}$, we take the charge in the segment of the circle to be $(3,1)$. Then, using \eqref{eq:change} the change of the charge is given as
\begin{align}
&(3,1) \overset{A}{\longrightarrow} (2,1) \overset{A}{\longrightarrow} \cdots \overset{A}{\longrightarrow} (-3,1) \overset{B}{\longrightarrow} (-1,-1) \overset{B}{\longrightarrow} (1,-3) \label{eq:change5} \\ & \overset{B}{\longrightarrow} (3,-5) \nonumber  \overset{N}{\longrightarrow} (3,-2) \overset{N}{\longrightarrow} (3,1), 
\end{align}
where the symbols on top of the arrows represents the fact that 5-brane crosses the cut emanating from the 7-brane of the corresponding type. The 5-brane charge goes back to the initial value $(3,1)$, as already mentioned. 

Then, we pull out the 7-branes from the inside of the circle along the cut. The formula \eqref{eq:monod} and the change in the 5-brane charge \eqref{eq:change5} give the number of 5-branes created by Hanany-Witten effect when the 7-brane crosses the circle of 5-brane. We have one extra $(1,0)$ 5-brane attached to $\mathbf{A}$, extra two $(1,-1)$ 5-branes attached to $\mathbf{B}$, and extra three $(0,1)$ 5-branes attached to $\mathbf{N}$ respectively after crossing a circle of 5-brane. 

\begin{figure}
	\centering
	\includegraphics[width=0.7\linewidth]{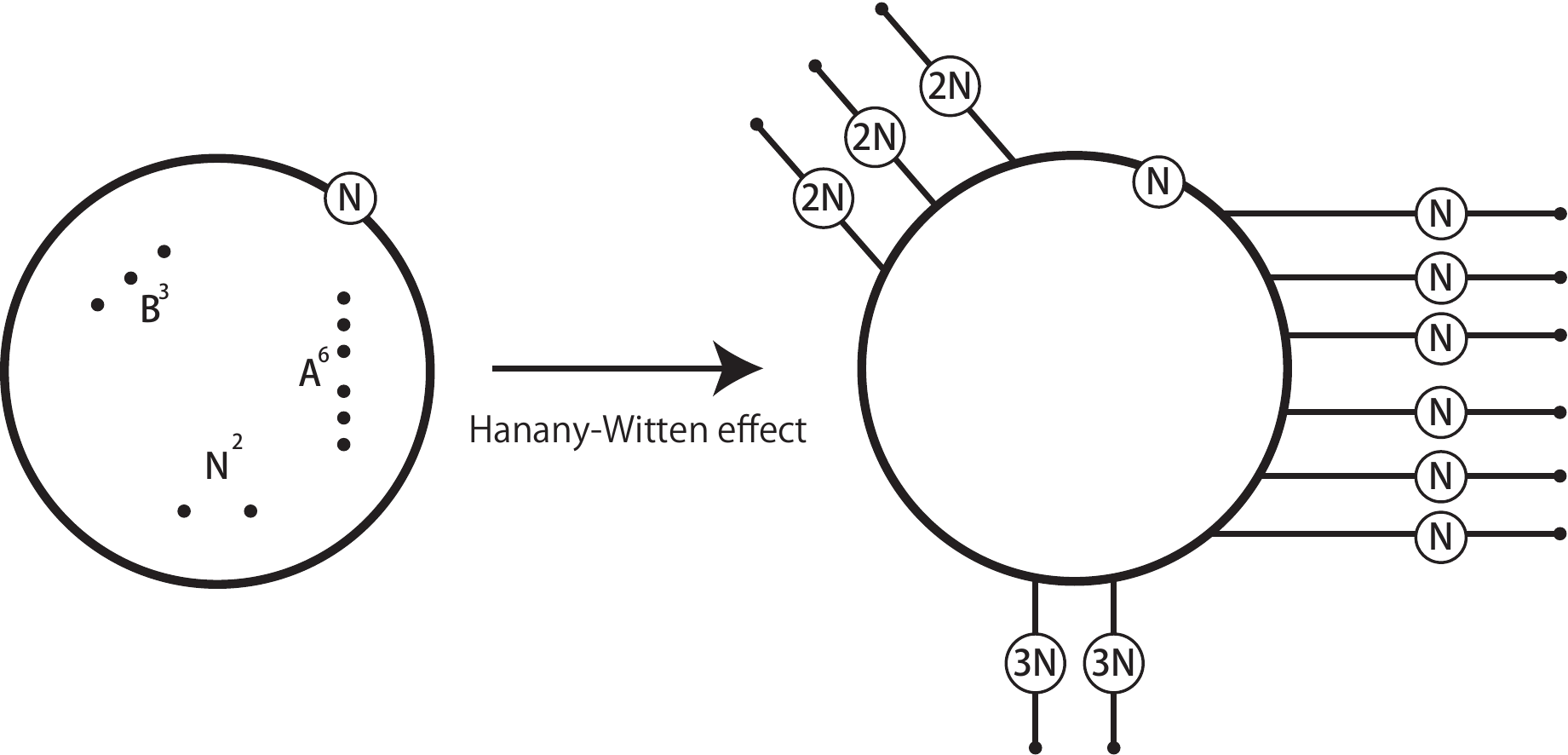}
	\caption{Pulling out eleven 7-branes $\mathbf{A}^6\mathbf{B}^3 \mathbf{N}^2$ from the inside of the $N$ circles of 5-brane creates the 5-brane junction with three legs due to Hanany-Witten effect. Each leg consists of $6N$ 5-branes. These 5-branes are grouped as shown in the right hand side of the figure and each group is terminated at a 7-brane.}
	\label{fig:BBT2}
\end{figure}

Finally, we have a three-pronged junction of 5-branes where each legs have $6N$ 5-branes terminated at 7-branes as shown in Fig \ref{fig:BBT2}. The patterns of terminations correspond to the Young diagrams $Y_1=[N^6]$, $Y_2=[2N,2N,2N]$ and $Y_3=[3N,3N]$. For example, $N$ $(1,0)$ 5-branes are grouped into a bunch and are terminated at a single $\mathbf{A}$. 

This 5-brane web describes the 5d theory $\widehat{\mathsf{T}}_{K}\{Y_1,Y_2,Y_3\}$ \cite{Benini:2009gi}. Thus we have shown using T-duality and Hanany-Witten effect that the $S^1$ compactification of rank-$N$ E-string theory is the 5d uplift of the class S theory. 

\subsection{T-dual of 6d theory $\mathcal{T}^{6d}_{(*)}\{u_i\}$}\label{sec:genweb}

\begin{figure}
	\centering
	\includegraphics[width=0.6\linewidth]{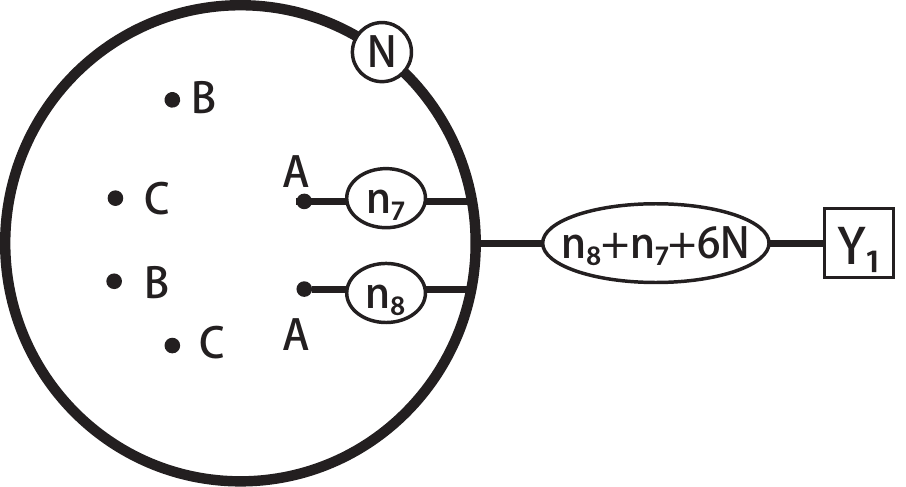}
	\caption{The Type IIB web for the 6d theory $\mathcal{T}^{6d}\{u_i\}$ on $S^1$ with Kaluza-Klein modes. We have $N$ circles of 5-branes. Outside the circles, we have a leg of $6N+n_7+n_8$ 5-branes terminated at 7-branes as specified by the partition $Y_1$. Inside the circle, we have six 7-branes $\mathbf{A}^2 \mathbf{B} \mathbf{C} \mathbf{B} \mathbf{C}$. $n_7$ and $n_8$ 5-branes are attached to the two $\mathbf{A}$ 7-branes respectively.}
	\label{fig:AABCBCloop}
\end{figure}

We generalize the result of Sec \ref{sec:BBT} to $\mathcal{T}^{6d}\{u_i\}$ in this subsection. To this end, we T-dualize the Type I' brane configuration in lower half of Figure \ref{fig:HZHW}. The resulting Type IIB configuration is illustrated in Figure \ref{fig:AABCBCloop}. We note that the case considered in Sec \ref{sec:BBT} corresponds to $n_7=n_8=0$ and $Y_1=[N^6]$.

The O8${}^-$ plane and two D8 branes at $x^6=0$ become six 7-branes $\widehat{\mathbf{E}}_3=\mathbf{A}^2\mathbf{B}\mathbf{C}\mathbf{B}\mathbf{C}$. The NS5 branes become the $N$ circles of 5-branes wrapping the six 7-branes $\widehat{\mathbf{E}}_3=\mathbf{A}^2\mathbf{B}\mathbf{C}\mathbf{B}\mathbf{C}$. We also have D6 branes in the Type I' configuration, which become extra (1,0) 5-branes in the Type IIB setup. $n_7$ and $n_8$ (1,0) 5-branes are attached to two $\mathbf{A}$ 7-branes wrapped by the $N$ circles of 5-branes respectively. These extra 5-branes extend toward the infinity and we have $6N+n_7+n_8$ 5-branes out of the circles due to Hanany-Witten effect. They are terminated at $\mathbf{A}$ type 7-branes, which come from $6+u_N$ D8 branes sitting where $x^6$ is very large in the Type I' configuration. The ending pattern is specified by the Young diagram $Y_1$ in \eqref{eq:Y1def}. 

The setup in Figure \ref{fig:AABCBCloop} includes the Kaluza-Klein modes. The decoupling of these modes can be done as in Sec \ref{sec:BBT} by rewriting $\widehat{\mathbf{E}}_3=\mathbf{E}_3 \mathbf{X}_{[3,1]}$ and moving $\mathbf{X}_{[3,1]}$ toward the infinity. Again, no additional 5-branes are created during the decoupling and we have five 7-branes $\mathbf{E}_3=\mathbf{A}^2\mathbf{B}\mathbf{C}\mathbf{C}$ inside the circles.


\begin{figure}
\begin{tabular}{ccc} 
\begin{minipage}{0.33\linewidth}
\begin{center}
\includegraphics[width=0.9\linewidth]{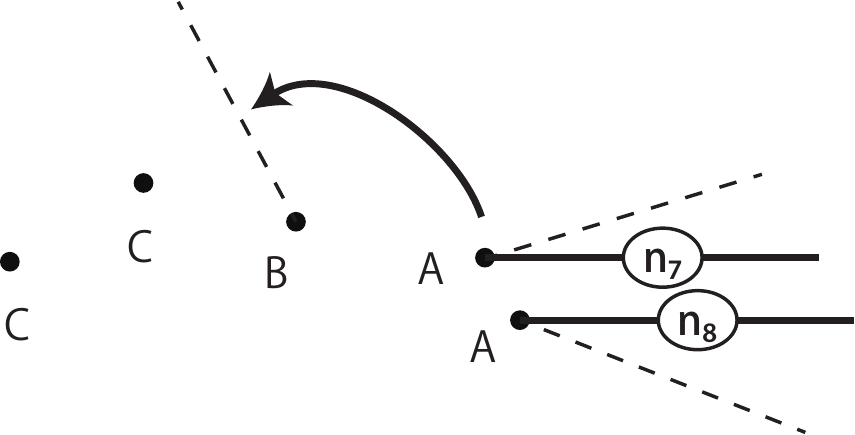}
\end{center}
\end{minipage}
\begin{minipage}{0.33\linewidth}
\begin{center}
\includegraphics[width=0.9\linewidth]{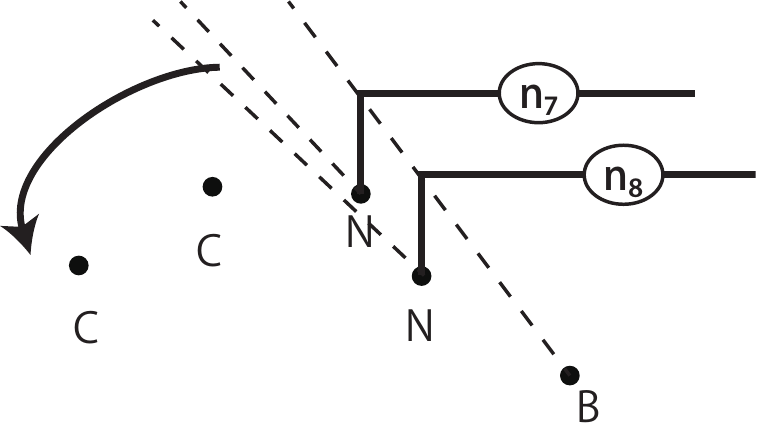}
\end{center}
\end{minipage}
\begin{minipage}{0.33\linewidth}
\begin{center}
\includegraphics[width=0.9\linewidth]{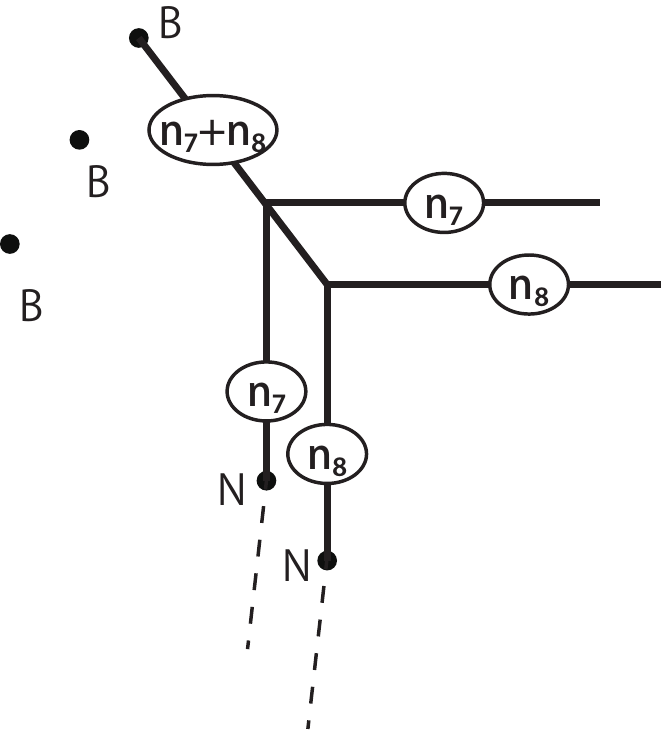}
\end{center}
\end{minipage}
\end{tabular}
\caption{The 7-brane rearrangement inside the circle of 5-branes. Extra $n_7$ and $n_8$ 5-branes attached to two $\mathbf{A}$ create the junction of 5-branes due to the Hanany-Witten effect. First, we move two $\mathbf{A}$ across the cut of $\mathbf{B}$. $\mathbf{A}$ becomes $\mathbf{N}$ and we obtain the middle configuration. Second, we move two $\mathbf{C}$s through the branch cuts of $\mathbf{N}$s. After that process, $\mathbf{C}^2$ becomes $\mathbf{B}^2$ since they cross the cuts from two $\mathbf{N}$. Finally, by moving one $\mathbf{B}$ along its cut, we obtain the configuration in right.}
\label{fig:AABCC}
\end{figure}

\paragraph{Pulling out 7-branes.} In order to obtain the 5-brane web as in Figure \ref{fig:5dfixture}, we rearrange the 7-branes inside the circles and pull them out toward the infinity. The rearrangement can be done by moving the 7-branes as in \eqref{eq:rearrange}. We carefully keep track the effect from the extra $n_7$ and $n_8$ $(1,0)$ 5-branes attached to the two $\mathbf{A}$ type 7-branes in Figure \ref{fig:AABCC}. After the rearrangement, one of the three $\mathbf{B}$s has new $n_7 + n_8$ 5-branes and the two $\mathbf{N}$s have new $n_7$ and $n_8$ 5-branes attached to it respectively.



\begin{figure}
	\centering
	\includegraphics[width=\linewidth]{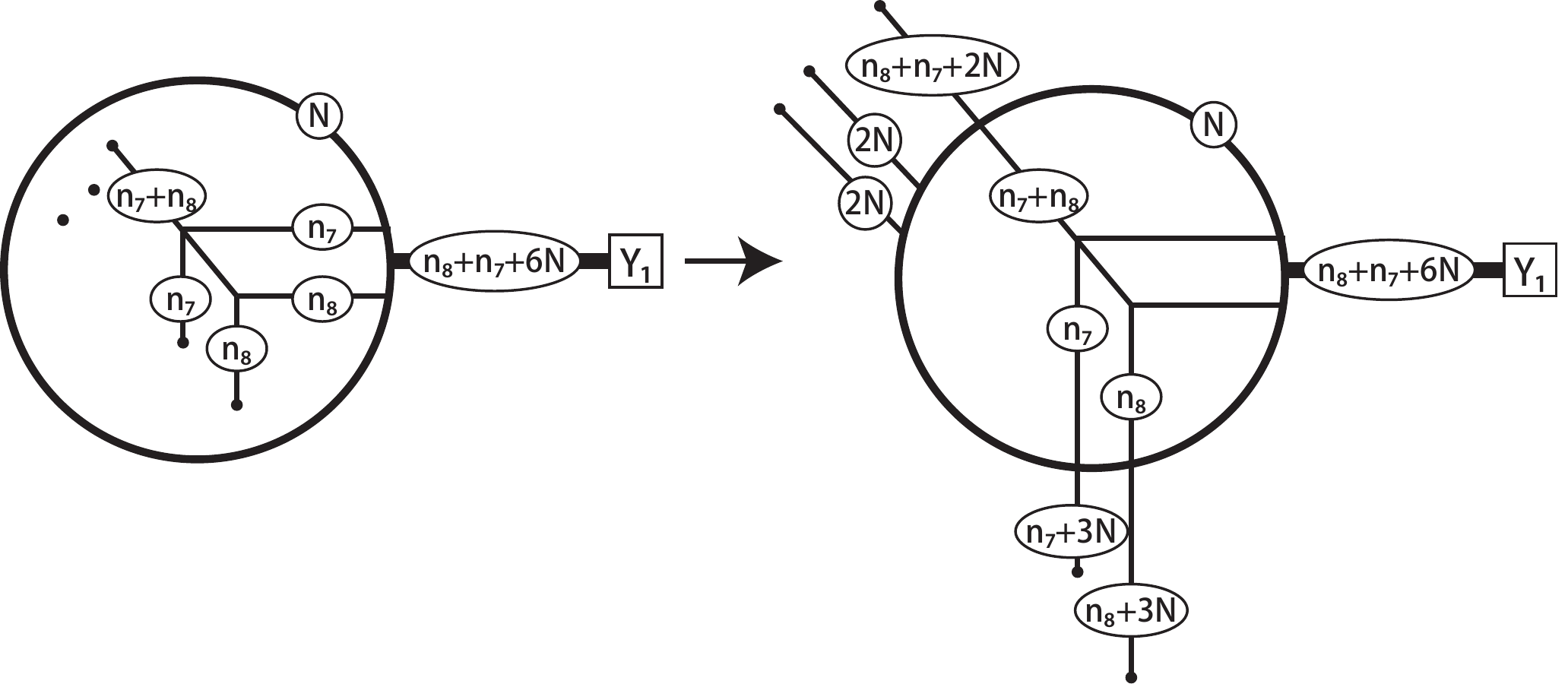}
	\caption{Pulling the eleven 7-branes from the inside of the circles of 5-branes, we again obtain the junction of 5-branes with three external legs. }
	\label{fig:AABCCloop2}
\end{figure}

Then, we pull all the 7-branes out of the circles. As in Sec \ref{sec:BBT}, we have one extra $(1,0)$ 5-brane attached to $\mathbf{A}$, extra two $(1,-1)$ 5-branes attached to $\mathbf{B}$, and extra three $(0,1)$ 5-branes attached to $\mathbf{N}$ respectively after crossing a circle of 5-brane. 
The result is shown in Figure \ref{fig:AABCCloop2}. We again have a three-pronged junction of 5-branes where each leg has $K=6N+n_7+n_8$ 5-branes terminated at 7-branes. The patterns of terminations are given by the Young diagrams $Y_1$, $Y_2=[2N+n_7+n_8, 2N,2N]$ and $Y_3=[3N+n_7, 3N+n_8]$. 

This is the 5-brane web which describes the 5d uplift $\widehat{\mathsf{T}}_{K}\{Y_1,Y_2,Y_3\}$ of the class S theory $\mathsf{T}_{K}\{Y_1,Y_2,Y_3\}$. Thus we have shown \eqref{eq:result1} using T-duality and Hanany-Witten effect. 

\begin{figure}
	\centering
	\includegraphics[width=0.6\linewidth]{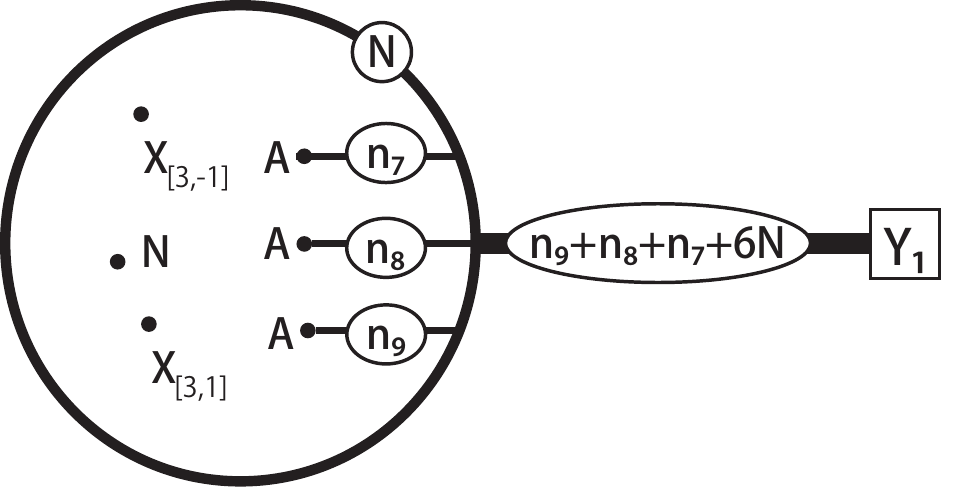}
	\caption{The Type IIB web for the 6d theory $\mathcal{T}_{*}^{6d}\{u_i\}$ on $S^1$ with Kaluza-Klein modes. We have $N$ circles of 5-brane. Outside the circles, we have a leg of $6N+n_7+n_8+n_9$ 5-branes terminated at 7-branes. Inside the circles, we have six 7-branes $\mathbf{A}^3 \mathbf{X}_{[3,-1]} \mathbf{N} \mathbf{X}_{[3,1]}$. $n_7$, $n_8$ and $n_9$ 5-branes are attached to three $\mathbf{A}$ 7-branes respectively.}
	\label{fig:AAAXNloop}
\end{figure}

\begin{figure}
\begin{tabular}{ccccc}
\begin{minipage}{0.33\linewidth}
\begin{center}
\includegraphics[width=\linewidth]{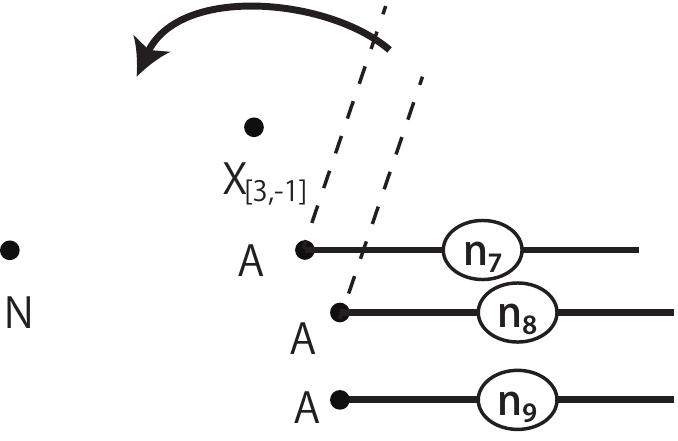}
\end{center}
\end{minipage}
\begin{minipage}{0.33\linewidth}
\begin{center}
\includegraphics[width=\linewidth]{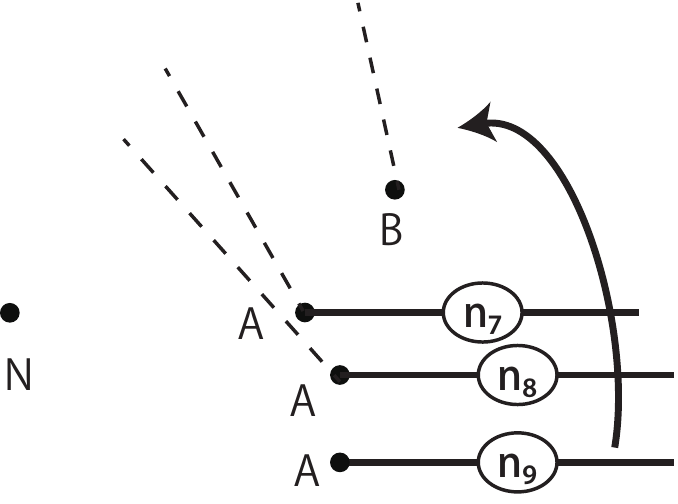}
\end{center}
\end{minipage}
\begin{minipage}{0.33\linewidth}
\begin{center}
\includegraphics[width=\linewidth]{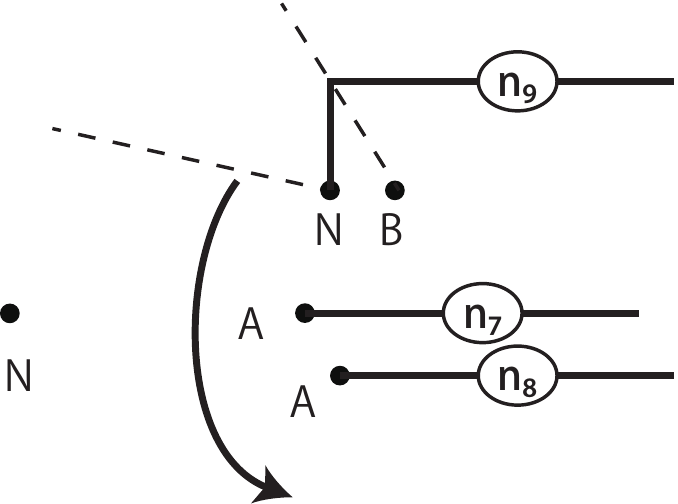}
\end{center}
\end{minipage}
\\
\\
\begin{minipage}{0.33\linewidth}
\begin{center}
\includegraphics[width=\linewidth]{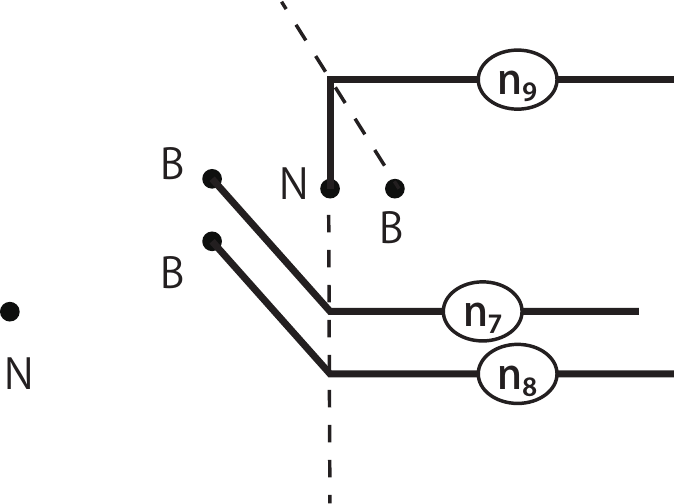}
\end{center}
\end{minipage}
\begin{minipage}{0.33\linewidth}
\begin{center}
\includegraphics[width=0.8\linewidth]{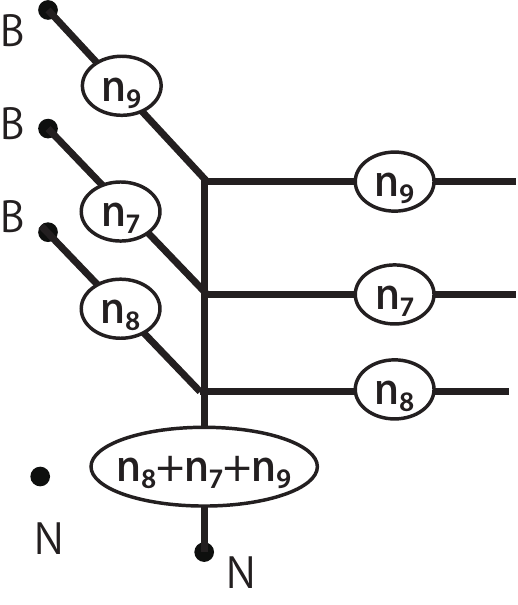}
\end{center}
\end{minipage}
\end{tabular}
\caption{The 7-brane rearrangement inside the circle of 5-branes. Extra 5-branes attached to the three $\mathbf{A}$ branes create the junction of 5-branes due to the Hanany-Witten effect.}
\label{fig:AAAXN}
\end{figure}

\paragraph{Case with O8${}^*$ plane.} 
Next we consider the $S^1$ compactification of the 6d theory $\mathcal{T}^{6d}_{*}\{u_i\}$ whose Type I' brane engineering uses the O8${}^*$ plane. To begin with, let us consider the T-dual of the O8${}^*$ plane.
As in Eq.~\eqref{eq:O8star}, the O8${}^*$ can be obtained by pulling two D8 branes from O8${}^-$+D8.
Noting that the T-dual of O8${}^-$+D8 is $\widehat{\mathbf{E}}_2$, the operation corresponding to (O8${}^-$+D8 $\rightarrow$ O8${}^*$, 2D8) in the Type IIB frame should be
\begin{equation}
	\widehat{\mathbf{E}}_2 = \mathbf{A}\widehat{\tilde{\mathbf{E}}}_1=\mathbf{A}^2\widehat{\mathbf{E}}_0.
\end{equation}
Therefore, we conclude that the T-dual of the O8${}^*$ plane is $\widehat{\mathbf{E}}_0$.

It is now straightforward to take T-dual of the 6d theory $\mathcal{T}^{6d}_* \{u_i\}$. The configuration is illustrated in Fig \ref{fig:AAAXNloop}. There are $N$ circles of 5-brane and there is a leg of $6N+n_7+n_8+n_9$ 5-branes outside the circles. The six 7-branes inside the circles are now $\mathbf{A}^3 \widehat{\mathbf{E}}_0$ where $\widehat{\mathbf{E}}_0=\mathbf{X}_{[3,-1]}\mathbf{N} \mathbf{X}_{[3,1]}$.

The decoupling of Kaluza-Klein modes can be done by moving $\mathbf{X}_{[3,1]}$ toward the infinity. Again, no additional 5-branes are created during the decoupling and we have $\mathbf{A}\mathbf{A}\mathbf{A} \mathbf{E}_0$ where $\mathbf{E}_0 = \mathbf{X}_{[3,-1]}\mathbf{N}$ inside the circles. 

In order to obtain the 5-brane web as in Figure \ref{fig:5dfixture}, we rearrange the 7-branes and pull them out from the circles. The required rearrangement is given as
\begin{equation}
\mathbf{A} \mathbf{A}^2 \mathbf{X}_{[3,-1]} \mathbf{N} = \mathbf{A} \mathbf{B} \mathbf{A}^2 \mathbf{N} = \mathbf{B} \mathbf{N} \mathbf{A}^2 \mathbf{N} = \mathbf{B}^3\mathbf{N}^2. \label{eq:change'}
\end{equation}
Taking account for the fact that there are extra $n_{7,8,9}$ 5-branes attached to the three $\mathbf{A}$s in Eq.~\eqref{eq:change'}, the brane rearrangement is illustrated in Fig \ref{fig:AAAXN}.

 By pulling all the 7-branes out of the circles, we again have a three-pronged junction of 5-branes where each leg has $K_*=6N+n_7+n_8+n_9$ 5-branes. Now we have three Young diagrams $Y_1$, $Y_2^*=[2N+n_7, 2N+n_8, 2N+n_9]$ and $Y_3^*=[3N+n_7+n_8+n_9, 3N]$. Therefore, we have shown the result \eqref{eq:result2}.

\section{4d conformal anomalies}\label{sec:anomalies}
In this section we compute the conformal and flavor central charges for the 4d theories $\cT^\text{4d}\{u_i\}$ and $\mathsf{T}_{K}\{Y_1,Y_2,Y_3\}$, and find the agreement. This provides another evidence for our claims \eqref{eq:claim4d} and \eqref{eq:claim4d2}.

In this section we assume $u_i\ge 1$ for $i=2, \cdots ,N$. Otherwise, the 6d theory is the higher rank E-string theory and the agreement of the central charges was already checked in \cite{Ohmori:2015pua, Benini:2009gi}.

\subsection{Central charges of $\cT^\text{4d}_{(*)}\{u_i\}$ from 6d anomaly polynomial}\label{sec:anom}
The conformal anomalies $a$, $c$ and the flavor central charge $k_i$ for the flavor symmetry $\mathfrak{f}_i$ were calculated in \cite{Ohmori:2015pua} for the 4d \Nequals{2} theory $\cT^\text{4d}\{u_i\}$. 
They are given as
\begin{equation}
	a= 24\alpha-12\beta-18\gamma,\qquad
	c= 64\alpha-12\beta-8\gamma,\qquad
	k_i=48\sigma_i,
\end{equation}
where $\alpha,\beta,\gamma$ and $\sigma_i$ are the coefficients of the anomaly polynomial 8-form $I^\text{6d}$ of the 6d theory $\cT^\text{6d}\{u_i\}$, defined by\footnote{Our normalizations for central charges and anomaly polynomial are those of \cite{Ohmori:2015pua,Ohmori:2015pia}}
\begin{align}
	I^\text{6d} \supset \alpha p_1(T)^2+\beta p_1(T)c_2(F_R) +\gamma p_2(T) +\sum_i \sigma_i p_1(T)c_2(F_{\mathfrak{f}_i}).
\end{align}
Here, $p_i(T)$ is the $i$th Pontryagin class of the tangent bundle and $c_2(F)=\frac14\mathrm{Tr}F^2$ is the second Chern class of the $R$- or flavor symmetry bundle, where $F_{\mathfrak{f}_i}$ is the background field strength for the global symmetry $\mathfrak{f}_i$.
It is convenient to define the effective numbers $n_v$ and $n_h$ of vector and hyper multiplets by
\begin{align}
	n_v= 8a -4c=-16(4\alpha+3\beta+7\gamma),\qquad n_h= 20c-16a=16(56\alpha-3\beta+8\gamma).
\end{align}

The algorithm for computing $I^\text{6d}$ was provided in \cite{Ohmori:2014kda}.
The anomaly polynomial $I^\text{6d}$ splits into two parts as
\begin{equation}
	I^\text{6d}=I^\text{one-loop}+I^\text{GS},
\end{equation}
where $I^\text{one-loop}$ is the naive one-loop contribution from the massless matter contents at a generic point on the tensor branch. $I^\text{GS}$ is the contribution from the 6d Green-Schwarz term given by
\begin{equation}
	I^\text{GS}= \frac{1}{2}\eta^{ij}I_i I_j,
\end{equation}
where $I_i$ are 4-forms topologically coupled to the self-dual two forms $B_i$ by the action 
\begin{equation}
	\eta^{ij}\int B_iI_j.
\end{equation}
Here $\eta^{ij}$ is the kinetic matrix in the effective Lagrangian for the tensor multiplet scalars $a_i$
and the gauge field strengths $F_{\fg_i}$;
\begin{equation}
	2\pi\int \eta^{ij}\left( \frac14 a_i \mathrm{Tr}F_j\wedge \star F_j -\frac12 \mathrm{d}a_i\wedge \star\mathrm{d}a_j\right).
\end{equation}
Here we also give the tensor vev to the scalar operator $a_1$ of the E-string theory in Figure~\ref{fig:quiver}.
For our case, $\eta^{ij}$ is determined to be
\begin{equation}
	\eta^{ij}=
	\begin{pmatrix}
		1&-1&&&\\
		-1&2&-1&&\\
		  &-1&2&-1&\\
		  &&&\ddots&-1\\
		  &&&-1&2
	\end{pmatrix}
\end{equation}
by the F-theory construction \cite{Heckman:2013pva,DelZotto:2014hpa} or the anomaly cancellation.

Using the formulas in \cite{Sadov:1996zm,Ohmori:2014kda}, we can determine the Green-Schwarz coupling $I_i$ and the kinematic matrix $\eta^{ij}$ for the 6d theory $\cT^\text{6d}_{(*)}\{u_i\}$, which is given as
\begin{equation}
	I^i= \eta^{ij}I_j = \eta^{ij}c_2(F_{\mathfrak{g}_j}) - \frac{1}{4}K^i p_1(T) +h^\vee(\mathfrak{g}_i)c_2(F_R)-c_2(F_{\mathfrak{f}_i}).
\end{equation}
In our case, $K^i=2-\eta^{ii}$ is given as $K^1=1, K^i=0 \; (i\geq2)$ and $h^\vee (\mathfrak{g}_i)$ is $h^\vee(\mathfrak{g}_1)=1, h^\vee(\mathfrak{g}_i)=h^\vee(\mathfrak{su}(u_i))=u_i \;(i\geq2)$.


Then the relevant part of the Green-Schwarz contribution $I^\text{GS}$ is 
\begin{equation}
	\begin{split}
		I^\text{GS} &\supset \frac{1}{32} \eta_{ij}K^i K^jp_1(T)^2 -\frac14 \eta_{ij}K^ih^\vee(\mathfrak{g}_j) p_1(T)c_2(F_R) + \frac14\eta_{ij}K^ic_2(F_{\mathfrak{f}_j})\\
				   &= \frac{N}{32}p_1(T)^2-\frac14 \left(N+\sum_{i=2}^{N}(N+1-i) u_i \right)p_1(T)c_2(F_R)\\
		     &\hspace{170pt}+\frac14\sum_{i=1}^{N} (N+1-i) p_1(T)c_2(F_{\mathfrak{f}_i}).
	\end{split}
\end{equation}
Here we have used the explicit form of the inverse $\eta_{ij}$ of the matrix $\eta^{ij}$;
\begin{equation}
	\eta_{ij}=
	\begin{pmatrix}
		N&N-1&N-2&\cdots&1\\
		N-1&N-1&N-2&\cdots&1\\
		N-2&N-2&N-2&\cdots&1\\
		\vdots&\vdots&\vdots&\ddots&\vdots\\
		1&1&1&\cdots&1
	\end{pmatrix}.
\end{equation}
Therefore, the Green-Schwarz contribution to the 4d conformal anomalies are
\begin{align}
	\delta n_v &= -2N +12\left(N+\sum_{i=2}^{N}(N+1-i) u_i \right),\\
	\delta n_h &= 28N + 12\left(N+\sum_{i=2}^{N}(N+1-i) u_i \right),\\
	\delta k_i &= 12(N+1-i).
\end{align}
Adding the contribution from the massless multiplets, the total 4d conformal anomalies are
\begin{align}
	n_v &= 11N+\sum_{i=2}^{N}\left( u_i^2-1 +12(N+1-i)u_i\right),\label{eq:nv}\\
	n_h &= 40N+\sum_{i=2}^{N}\left( 2u_i^2+ 12(N+1-i)u_i \right)-\sum_{i=2}^{N-1}u_iu_{i+1},\label{eq:nh}\\
	k_i &= 12(N+1-i)+2u_i \quad\quad ( i=1,\cdots,N).
\end{align}

Additionally, the complex dimension of the Coulomb branch of $\cT^\text{4d}\{u_i\}$ is just the sum of the number of 6d tensors and the ranks of the gauge groups;
\begin{equation}
	\mathrm{dim}_{\mathbb{C}}\mathrm{Coulomb}= \sum_{i=2}^N (u_i-1) +N =1+\sum_{i=2}^{N}u_i.
	\label{eq:dimCoulomb}
\end{equation}

\subsection{Central charges of $\mathsf{T}_{K}\{Y_1,Y_2,Y_3\}$ from class S formulas}\label{sec:classS}
In this subsection we calculate the conformal anomalies of the class S theory $\mathsf{T}_{K}\{Y_1,Y_2,Y_3\}$. First, we briefly recall the central charge formulas in \cite{Chacaltana:2010ks,Chacaltana:2012zy}.

Let $Y^\text{T}=[\ell_1,\cdots,\ell_N]$ be the partition of $K$ obtained by taking the transpose of the Young diagram $Y$.
The pole structure $\{p_k\}$, $k=1,\cdots,Y-m$ of $Y$ is defined by
\begin{equation}
	\begin{cases}
		p_1=0,\\
		p_{k+1}-p_{k}=0 &\text{if $k$ is equal to $\ell_i$ for some $i$},\\
		p_{k+1}-p_{k}=1 &\text{otherwise},
	\end{cases}
\end{equation}
therefore it looks like
\begin{align}
	\{p_k\}=\{0,1,2\cdots ,\ell_1-1,\ell_1-1,\ell_1,\cdots,\ell_1+\ell_2-2,\cdots, K-m\}.
\end{align}

For the class S theory $\mathsf{T}_{K}\{Y_1,Y_2,Y_3\}$, the number $d_k$ of the Coulomb branch operators with dimension $k$ is given as
\footnote{The formulas below are valid only when $\sum_ip^{(i)}_k\ge 2k-1$. When $u_i=0$ which corresponds to the higher rank $E_8$ Minahan-Nemeschansky theory, the pole structure for the class S description violates this bound. That case was studied well in \cite{Benini:2009gi} as already mentioned.}
\begin{equation}
	d_k=1-2k+\sum_{i=1}^{3} p^{(i)}_k\label{eq:Cdim}
\end{equation}
where $\{p^{(i)}_k\}$ is the pole structure of $Y_i$. The effective number of vectors $n_v$ is
\begin{equation}
	n_v=\sum_{k=2}^{K} (2k-1)d_k, \label{eq:vnumber}
\end{equation}
and the formula for $n_h$ is
\begin{align}
	n_h &=-\frac43(K^3-K)+\sum_{n=1}^3 f(Y_n),\label{eq:hnumber} \\
	f(Y) &=\frac12\left(-K+\sum_i \ell_i^2\right)+\sum_{k=2}^{K}(2k-1)p_k.
\end{align}

Let us apply the formulas \eqref{eq:Cdim}, \eqref{eq:vnumber} and \eqref{eq:hnumber} to the class S theory $\mathsf{T}_{K}\{Y_1,Y_2,Y_3\}$ where $K=6N+n_7+n_8$, $Y_1$ is defined by \eqref{eq:Y1def}, $Y_2=[2N+n_7+n_8,2N,2N]$ and $Y_3=[3N+n_7,3N+n_8]$. After some calculation, we obtain
\begin{align}
	&n_v = 10N+1+\sum_{i=2}^{N}(u_i^2+12(N+1-i)u_i),
	\label{eq:Snv} \\
	&n_h = 40N+\sum_{i=2}^{N}\left( 2u_i^2+ 12(N+1-i)u_i \right)-\sum_{i=2}^{N-1}u_iu_{i+1},
	\label{eq:Snh} \\
	&\mathrm{dim}_{\mathbb{C}}\mathrm{Coulomb} = \sum_{k=2}^{K}d_k =1+\sum_{i=2}^{N}u_i, 
\end{align}
which agree with the results \eqref{eq:nv}, \eqref{eq:nh} and \eqref{eq:dimCoulomb}. 

We can also check the agreement of flavor groups and their central charges. 
As explained in \cite{Chacaltana:2012zy}, the theory $\mathsf{T}_{K}\{Y_1,Y_2,Y_3\}$ has the flavor group (up to $\mathfrak{u}(1)$ factors)
\begin{equation}
	\fsu(\ell_1-\ell_2)_{2\ell_1}\times \fsu(\ell_2-\ell_3)_{2L_2}
	\times \cdots\times \fsu(\ell_{N-n_6})_{12N}
	\times\fsu(2)_{12N},
\end{equation}
where the subscripts denote the flavor central charges and $L_i$ is defined by $L_i=\sum_{j=1}^i \ell_j$.
There is an additional $\su(2)_{2K}$ when $n_7=n_8$, and moreover $\su(2)_{12N}$ enhances to $\su(3)_{12N}$ when $n_7=n_8=0$.
When $n_8\neq n_7\neq 0$, $\fsu(\ell_{N-i+1}-\ell_{N-i+2})_{2L_{N-i+1}}=\fsu(2u_i-u_{i+1}-u_{i-1})_{12(N+1-i)+2u_i}$ is nothing but the flavor group $\ff_i$ and its central charge of $\cT^\text{4d}\{u_i\}$,
and $\su(2)_{12N}$ should be identified with $\ff_1$.
One can also match the flavor groups and central charges for other cases.

In the discussion so far, we only considered the 4d theory $\mathsf{T}_{K}\{Y_1,Y_2,Y_3\}$. It is straightforward to compute those quantities for the 4d theory $\mathsf{T}_{K_*}\{Y_1,Y^*_2,Y^*_3\}$ and check the agreement with the results in Sec \ref{sec:anom}.

\section{Conclusions and discussions}\label{sec:conclusions}

In this paper, we have established the dualities \eqref{eq:result1}, \eqref{eq:claim4d}, \eqref{eq:result2} and \eqref{eq:claim4d2}.
We started from the Type I' construction of the 6d theory $\cT^\text{6d}_{(*)}\{u_i\}$. Then we took T-dual to obtain the 5d web and read off the class S construction from the web.
We also checked the class S description by comparing the 4d conformal anomalies computed from the 6d anomaly polynomial \cite{Ohmori:2015pua} and from the class S formulas \cite{Chacaltana:2010ks,Chacaltana:2012zy}.

Here we would like to list some future directions.

\paragraph{Other theories Higgsable to E-string theories with $\su$ and $\mathfrak{usp}$ groups.}
In this paper we concentrated on the case where the gauge group on the $-1$ curve is empty \eqref{eq:tbranch}.
As an immediate generalization, one can consider the theory like
\begin{equation}
	\begin{matrix}
		[\ff_1]& \fg_1 &\fsu(u_2)&\fsu(u_3)& \cdots &\fsu(u_N)\\
		& 1 & 2 &2&\cdots & 2 
	\end{matrix},\label{eq:-1gauge}
\end{equation}
where $\mathfrak{g}_1$ is $\mathfrak{su}(u_1)$ or $\mathfrak{usp}(2u_1)$ with some matters which cancel the gauge anomalies. Since the method to obtain $\su$ with an antisymmetric hyper or $\mathfrak{usp}$ from the 5-brane web was found in \cite{Bergman:2015dpa},
the compactification of such a 6d theory is also expected to be described by a 5-brane web.



\paragraph{Theories Higgsable to E-string theories with other groups.} 
One can further generalize \eqref{eq:-1gauge} to 
\begin{equation}
	\begin{matrix}
		[\ff_1]& \fg_1 &\fg_2&\fg_3& \cdots &\fg_N\\
		& 1 & 2 &2&\cdots & 2 
	\end{matrix},
\end{equation}
with general groups $\fg_i$. Between non-$\su$ groups, there are appropriate minimal conformal matters \cite{DelZotto:2014hpa}.
When $\fg_i$ are of $\mathfrak{so}$ type for sufficiently large $i$, the theory is expected to have the Type I' brane construction with an O6 plane along D6 branes, but T-duality introduces O5 planes in the 5-brane web. It is interesting to develop a way to treat O5 planes in the 5d web and investigate how these 5d theories are related to the class S theories of D-type.

\section*{Acknowledgments}
The authors thanks Masato Taki, Tom Rudelius, and Yuji Tachikawa for helpful discussions.
KO and HS are partially supported by the Programs for Leading Graduate Schools, MEXT, Japan,
via the Advanced Leading Graduate Course for Photon Science
and via the  Leading Graduate Course for Frontiers of Mathematical Sciences and Physics, respectively. 
KO is also supported by JSPS Research Fellowship for Young Scientists.

\appendix
 
\bibliographystyle{ytphys}
\bibliography{ref3}

\end{document}